\documentclass[conference]{IEEEtran}
\IEEEoverridecommandlockouts
\usepackage{cite}
\usepackage{amsmath,amssymb,amsfonts}
\usepackage{algorithmic}
\usepackage{graphicx}
\usepackage{textcomp}
\usepackage{multirow}
\usepackage{subcaption}
\usepackage[table,xcdraw]{xcolor}
\def\BibTeX{{\rm B\kern-.05em{\sc i\kern-.025em b}\kern-.08em
    T\kern-.1667em\lower.7ex\hbox{E}\kern-.125emX}}

\usepackage{booktabs} 

\begin{document}
\setlength{\textfloatsep}{3pt}
\title{\huge Noise Aware Utility Optimization of NISQ Devices}

\author{

 \IEEEauthorblockN{Jean-Baptiste Waring}
 \IEEEauthorblockA{\textit{\footnotesize Dept. of Electrical \& Computer Engineering} \\
 \textit{Concordia University}\\
 Montréal, Québec \\
 j\_warin@live.concordia.ca}
 \and
 \IEEEauthorblockN{Christophe Pere}
 \IEEEauthorblockA{\textit{\footnotesize Dept. of Computer Science \& Software Engineering} \\
 \textit{INTRIQ, Laval University,}\\
 Québec, QC, Canada \\
 PINQ2, \\
 christophe.pere.1@ulaval.ca}
 \and
 \IEEEauthorblockN{Sébastien Le Beux}
 \IEEEauthorblockA{\textit{\footnotesize Dept. of Electrical \& Computer Engineering} \\
 \textit{Concordia University}\\
 Montréal, Québec \\
 sebastien.lebeux@concordia.ca}
}

\maketitle

\begin{abstract}
In order to enter the era of utility, noisy intermediate-scale quantum (NISQ) devices need to enable long-range entanglement of large qubit chains. 
However, due to the limited connectivity of superconducting NISQ devices, long-range entangling gates are realized in linear depth.
Furthermore, a time-dependent degradation of the average CNOT gate fidelity is observed.
Likely due to aging, this phenomenon further degrades entanglement capabilities.
Our aim is to help in the current efforts to achieve utility and provide an opportunity to extend the utility lifespan of current devices ---albeit by selecting fewer, high quality resources. To achieve this, we provide a method to transform user-provided CNOT and readout error requirements into a compliant partition onto which circuits can be executed.
We demonstrate an improvement of up to 52\% in fidelity for a random CNOT chain of length 50 qubits and consistent improvements between 11.8\% and 47.7\% for chains between 10 and 40 in varying in increments of 10, respectively.
\end{abstract}

\vspace{-0.2cm}
\section{Introduction}
\IEEEPARstart{T}{he} maturation of quantum computing promises unprecedented computing power, capable of tackling problems that would remain intractable for classical computing paradigms. In recent years, the quantum computing landscape has witnessed significant advancements. However, as quantum computing hardware and algorithms grow in complexity, the need for efficient routing and circuit layout mechanisms is paramount.
The performance of quantum algorithms can be significantly impacted by the routing and layout techniques employed and their efficiencies. In architectures that only support near-neighbour connectivity, permutations of quantum states are necessary to satisfy coupling constraints because the no-cloning theorem \cite{wootters_single_1982} prohibits making a copy of a quantum state. This leads to a linear increase in circuit depth with the number of permutations, significantly constraining fidelity when running large circuits on NISQ\cite{preskill_quantum_2018} devices with low average qubit connectivity. Consequently, optimizing the outcomes of quantum computations necessitates a nuanced approach that not only acknowledges the general presence of noise but also adapts to the specific noise profiles and coupling constraints of individual devices.

\textbf{In this paper, we present an innovative approach to enhance expected gate fidelity on IBM's superconducting quantum devices.} By integrating a resource pruning strategy tailored to user-defined error thresholds, we dynamically adjust to individual quantum systems' unique and evolving noise profiles. Through extensive experimental validations, we demonstrate that our method significantly improves the fidelity of long-range CNOT gate operations. Our findings reveal that, in scenarios involving 20-qubit CNOT chains, our approach yields up to a 40\% increase in expected gate fidelity compared to no resource-pruning on a 127-qubit IBM Quantum Computer. This work addresses the challenges of performance degradations due to system aging but also paves the way for achieving higher computational accuracy in the NISQ-era on high-fidelity device partitions.

This paper is structured as follows: Section II gives some background on noise and its time evolution in NISQ devices. Section III outlines our resource-pruning methodology. Section IV presents and discusses the comparative results obtained through our methodology. Finally, Section V highlights future prospects and implications of our findings.
\vspace{-0.05cm}
\section{Background \& Related Work}

\begin{figure}[h!]
    \centering
    \includegraphics[width=\linewidth]{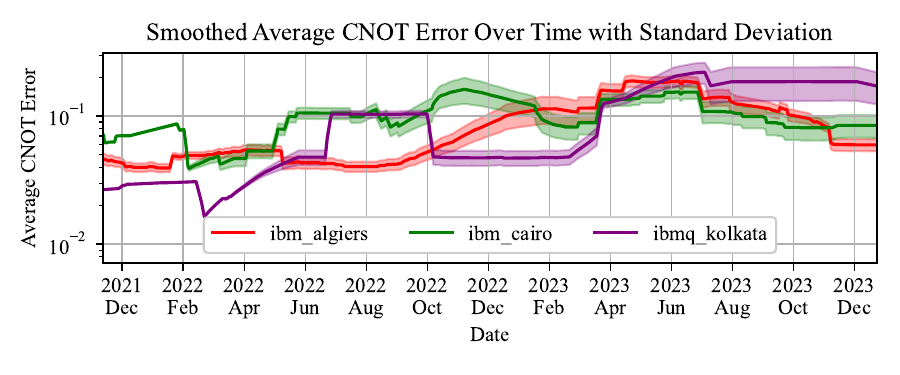}
    \caption{Smoothed Average CNOT Error Over Time with Standard Deviation, illustrating the temporal evolution of CNOT errors. Data collected using Qiskit  \cite{treinish_qiskitqiskit-metapackage_2023} in Jan. 24'. }
    \label{fig:degradation}
\end{figure}

\vspace{-0.15cm}
NISQ-era devices are highly affected by noise \cite{preskill_quantum_2018}\cite{de_micheli_advances_2022}. 
Furthermore, research by \cite{nachman_unfolding_2020} demonstrates that each IBM quantum computer has a unique, time-dependent noise signature that can be identified and learned over time. As depicted in Figure \ref{fig:degradation}, there is a discernible increase in the average CNOT error rate correlated with the age of a sample of devices from IBM's fleet. One contributing factor to this trend is the drift in the resistance \cite{burnett_decoherence_2019} of the Josephson Junctions used to create qubits as they age, resulting in qubit frequency fluctuations. We believe those fluctuations are responsible, in part, for the degradatation of gate fidelity on aging devices. Additional studies, including by \cite{schlor_correlating_2019} and \cite{hong_demonstration_2020}, independently confirm that T1 relaxation time variability of qubits is a primary contributor to the frequency and dephasing fluctuations, as well as to the observed variations in gate fidelity. These findings underscore the importance of addressing time-dependent qubit performance degradation for achieving reliable quantum computations. 

Adaptive noise-aware circuit compilation and transpilation techniques such as \cite{murali_noise-adaptive_2019}\cite{quetschlich_compiler_2023}\cite{ji_calibration-aware_2022} involve mapping quantum circuits to the physical qubit layout of a quantum processor while minimizing the impact of errors. By taking into account the specific error rates and connectivity of the quantum device, these techniques can significantly improve the fidelity of quantum computations. 
Our method provides a deterministic approach to selecting low-error resources prior to the transpilation stage, ensuring that the quantum circuits are mapped onto the most reliable parts of the quantum hardware. This pre-selection process complements adaptive noise-aware compilation techniques by establishing a high-fidelity partition, which can further be optimized by existing transpilation techniques.

\vspace{-0.1cm}
\section{Method}

The primary objective of our method is to increase the expected fidelity when executing quantum circuits on hardware with highly heterogeneous error characteristics. This is achieved by finding a hardware partition composed of qubits and connections that satisfies user-defined error criteria, namely: \textit{i)} \textbf{CNOT Error Threshold:} the maximum CNOT error rate for a connection between two qubits to be included to the partition, \textit{ii)} \textbf{Readout Error Threshold:} the maximum qubit readout error rate for a qubit to be included in the partition. Given those two parameters, our method provides a systematic and fully automated approach to obtain a partition of a quantum architecture that meets user-defined error criteria. It allows the mitigate the effect of resource heterogeneity as well as device degradation with time by providing the user with a smaller, low-error partition onto which smaller circuits can be mapped to extend the useful life of the device.
\begin{figure}[ht!]
    \centering
    \includegraphics[width=0.95\linewidth]{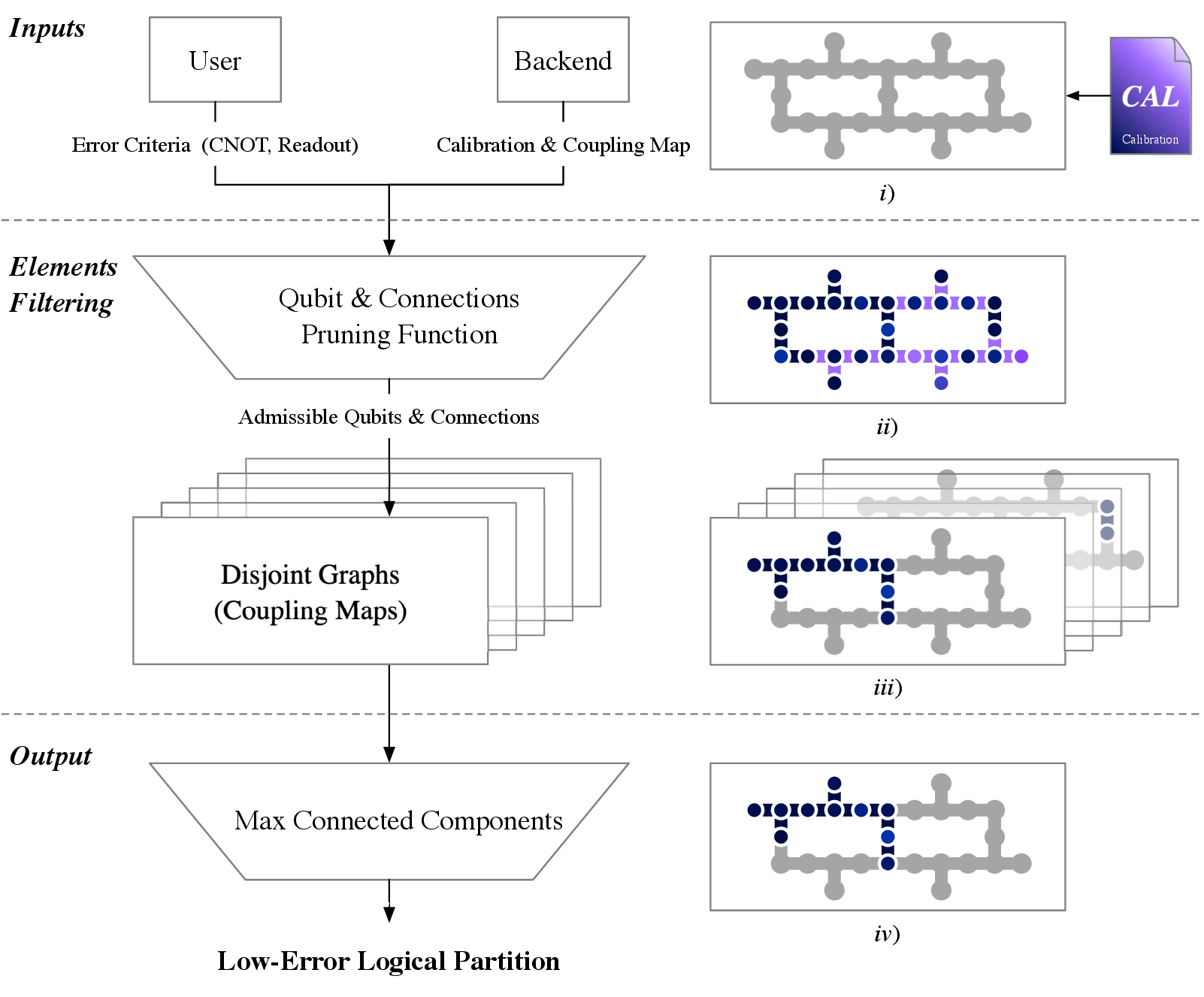}
    \caption{Our proposed method. \textit{i}) Coupling Map \& Calibration Data. \textit{ii}) Weighted Network Representation of the Device's Architecture. \textit{iii}) Disjoint Graphs. \textit{iv}) Partition that is returned to the user for circuit execution. Example from a 27-qubit IBM Quantum Computer.}
    \label{fig:flowchart}
\end{figure}
\vspace{-0.15cm}
\subsection{Inputs}
\vspace{-0.1cm}
Qiskit inherently recognizes and categorizes qubits with complete operational failure or significant performance issues as 'faulty' \cite{treinish_qiskitqiskit-metapackage_2023}. These qubits are flagged based on calibration data that suggests they are not meeting the minimum standards required for quantum computations. However, this binary classification—faulty or not—does not encapsulate the gradations of performance degradation that can occur over time as quantum hardware components age as well as between device calibrations.

\textbf{Our approach introduces a quantum resource pruning method through user-defined error thresholds. }Unlike the fixed definition of a faulty qubit, which may exclude only those qubits with extreme deviations in performance, user-defined thresholds allow for a more nuanced and adaptable utilization of the quantum processor's resources.
First, the coupling map --a directed acyclic graph or DAG-- of the target quantum device is obtained. In this graph, nodes symbolize the qubits, and edges denote the connections --or coupling-- between these qubits.
Then, we fetch the target quantum device's calibration data, allowing us to take into account the current error rates of qubits and their connections. We then use the calibration data to enrich the coupling map with error rates, thus converting it to a weighted network representation of the device's architecture. In this weighted network representation, the nodes (representing qubits) are assigned weights based on their readout error rates, and the edges (representing qubit couplings) reflect the CNOT error rates. It is important to highlight that modern IBM quantum devices use the Echoed Cross-Resonance (ECR) gate as their native entangling operation rather than the CNOT gate commonly found in theoretical quantum circuit designs \cite{jurcevic_demonstration_2021}\cite{sundaresan_reducing_2020}. The ECR gate, like the CNOT, creates entanglement but through a different mechanism suited to the hardware's architecture. Consequently, when we refer to the 'CNOT error rate' in this context, we are actually discussing the error rates associated with ECR operations. Given that quantum circuits are typically designed using CNOT gates and later transpiled to the device's native gate set, this terminology is used for simplicity. 

This integration of user-specified error thresholds with detailed device calibration and coupling constraints in weighted network form is the basis for the following quantum resource selection and optimization process.
\vspace{-0.2cm}
\subsection{Elements Pruning}
\vspace{-0.1cm}
With these parameters, our methodology employs a pruning function to sift through the qubits and connections based on their error rates.
First, it discards any elements whose error rates exceed the user-defined thresholds. Second, it removes disjoint elements that cannot compose a valid partition. For example, a lone edge cannot be admissible if the nodes it connects are not included.
After pruning, we are left with valid partitions that adhere to the error criteria. These partitions contain only the qubits and connections that meet the user-defined thresholds, ensuring that each element within them is within the bounds of acceptable error rates.
 \vspace{-0.2cm}
\subsection{Output}
\vspace{-0.1cm}
Finally, the largest partition is transformed into a valid Qiskit coupling map and returned to the user as the optimal partition. This largest partition not only informs the user about the maximum number of qubits available for their specific computational needs but also highlights the potential for parallel circuit execution. Indeed, smaller partitions, close in size to the largest, might allow for parallel execution of multiple copies of a circuit, optimizing resource use, enhancing computational throughput, and potentially diminishing overall runtime and cost for an experiment.

\vspace{-0.15cm}
\section{Results}
\vspace{-0.1cm}
\begin{figure*}[ht!]
    \centering
    \setlength\fboxsep{0pt}
    \setlength\fboxrule{0.0pt}

\fbox{\includegraphics[width=5.5in]{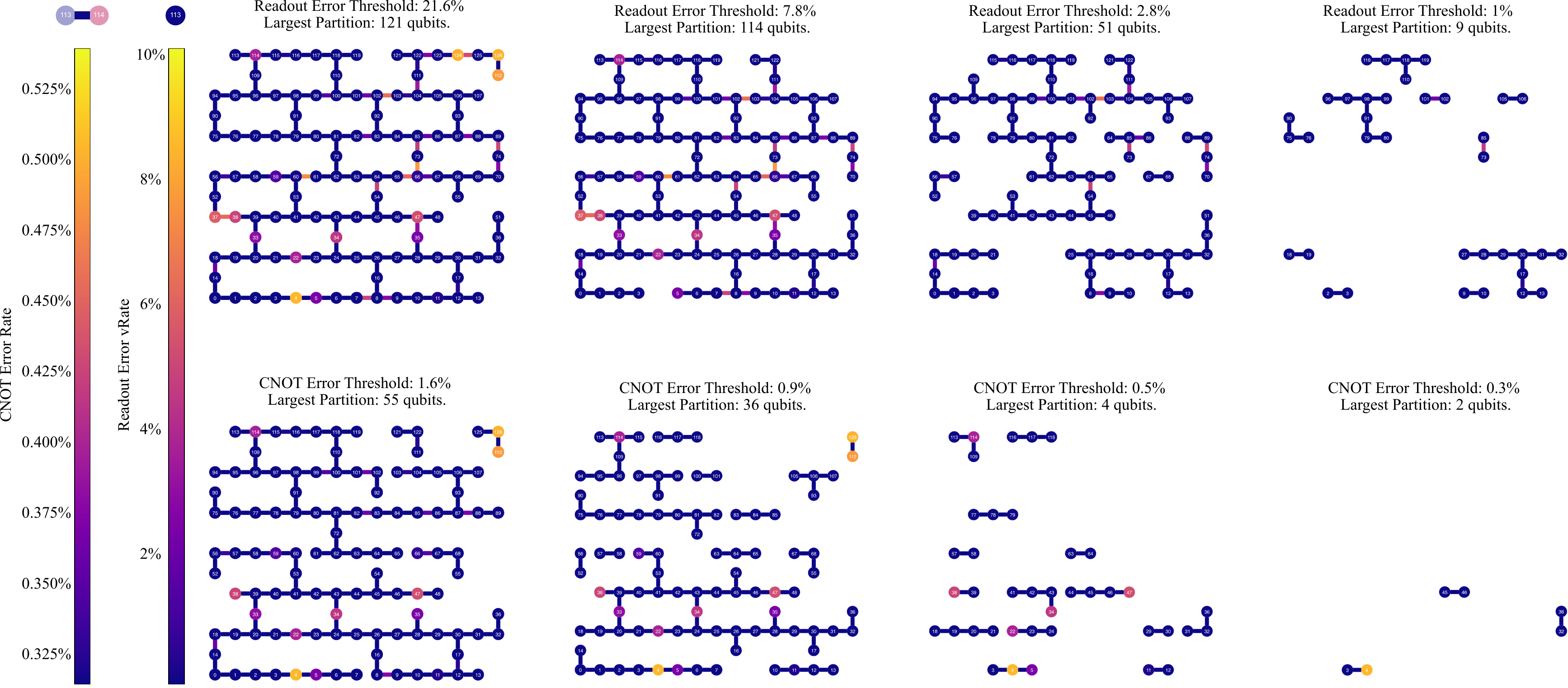}}

    \caption{Results of our method showing the impact of varying readout \& CNOT error thresholds on largest partition size. Calibration data from \texttt{ibm\_quebec} was obtained on January 30th, 2024. From left to right, the diagrams represent the largest partition extracted from a 127-qubit quantum processor's coupling map, corresponding to readout error thresholds ranging from 21.6\% to 1.0\% and CNOT error thresholds ranging from 1.6\% to 0.3\%. As the threshold decreases--indicating higher fidelity requirements--the size of the largest partition decreases, demonstrating the trade-off between lower error rates and the availability of computational resources.}
    \label{fig:readout-cnot-subgraphs}
\end{figure*}
In this section, we first showcase the outcome of our methodology as applied to the IBM Eagle 127-qubit Processor, specifically \texttt{ibm\_quebec}---a 127-qubit Eagle r3 IBM Quantum System One installed in Bromont, QC that entered service in September 2023. We proceed to evaluate our method's effectiveness through a gate fidelity benchmark using chains of 20 CNOTs. Finally, we extend our benchmark to assess the gate fidelity across varying lengths of CNOT chains and compare the results to a baseline established using the original, non-pruned partition.

\vspace{-0.1cm}
\subsection{Preliminary Analysis}
\vspace{-0.1cm}
In our analysis, we apply our methodology to \texttt{ibm\_quebec}. The resulting visualizations of the processor's coupling map—post-pruning—, for varying CNOT and Readout error thresholds are depicted in Figure \ref{fig:readout-cnot-subgraphs}. 
The result shows a complex yet intuitive trade-off dynamic between error thresholds and resulting partitions. As we impose more stringent fidelity requirements by lowering the acceptable error thresholds, the size of the largest partition decreases correspondingly. For instance, reducing the CNOT error threshold by approximately sixfold, from 1.6\% to 0.3\%, results in the largest partition shrinking from 55 qubits—nearly half of the machine's 127 qubits—to a mere 2, a 28-fold reduction. Additionally, as the error thresholds are lowered, there is a noticeable decrease in the average connectivity of the partitions, leading them to increasingly \textit{i)} resemble linear qubit chains \textit{ii)} retain very little to no features of the original heavy-hex topology. 
\vspace{-0.15cm}
\subsection{Benchmark Selection Rationale}
\vspace{-0.1cm}
To validate the effectiveness of our method, we present experimental results showing its impact on the gate fidelity of long-range CNOT gates. This serves as a benchmark for real-world improvements over the baseline. We choose CNOT chains as our benchmark due to their essential role in quantum circuits, specifically for facilitating the long-range entanglement required by algorithms like Shor's. Indeed, given the non-fully connected nature of IBM's superconducting qubits--where each qubit is not directly coupled to all others--, interactions between non-adjacent qubits necessitate chains--or cascades--of CNOT gates to move quantum information across the chip. This, in turn, enables the physical realization of sophisticated quantum algorithms designed with all-to-all connectivity in mind. Consequently, the fidelity of long-range CNOT gates is a critical factor in determining the utility of the quantum device.
\vspace{-0.15cm}
\subsection{Experimental Setup}
\vspace{-0.1cm}
Our experimental setup consists of generating random paths across the largest partitions obtained post-pruning, employing a random walk algorithm. Each of those paths will later lead to one circuit being executed on the system.
By invoking the \texttt{initial\_layout} flag, we direct Qiskit's transpiler to conform its transpilation process to the generated partitions, ensuring results integrity. Quantum Process Tomography (QPT) is used to estimate the fidelity of quantum gates by reconstructing the quantum process from measurements. This reconstruction involves comparing the theoretical and experimental process matrices, providing insight into how closely the quantum device performs to expectations. Once an estimate of the process fidelity is obtained, a simple linear relationship is used to convert it to gate fidelity: $F_{\textrm{gate}}=(4F_{\textrm{process}}+1 )/ 5$ \cite{horodecki_general_1999}\cite{nielsen_simple_2002}.
Gate fidelity is chosen because it provides a scalar value that encapsulates the overall performance of the quantum gate, making it a relevant figure of merit for our method's effectiveness.
Finally, the experiments were run over the course of three weeks on \texttt{ibm\_quebec}, totalling 3,000+ circuit executions of 4096 shots each. Overall, those experiments took about 492,000 Qiskit Runtime seconds--roughly 135 hours.
\begin{figure}[h!]
    \centering
    \includegraphics[width=\linewidth]{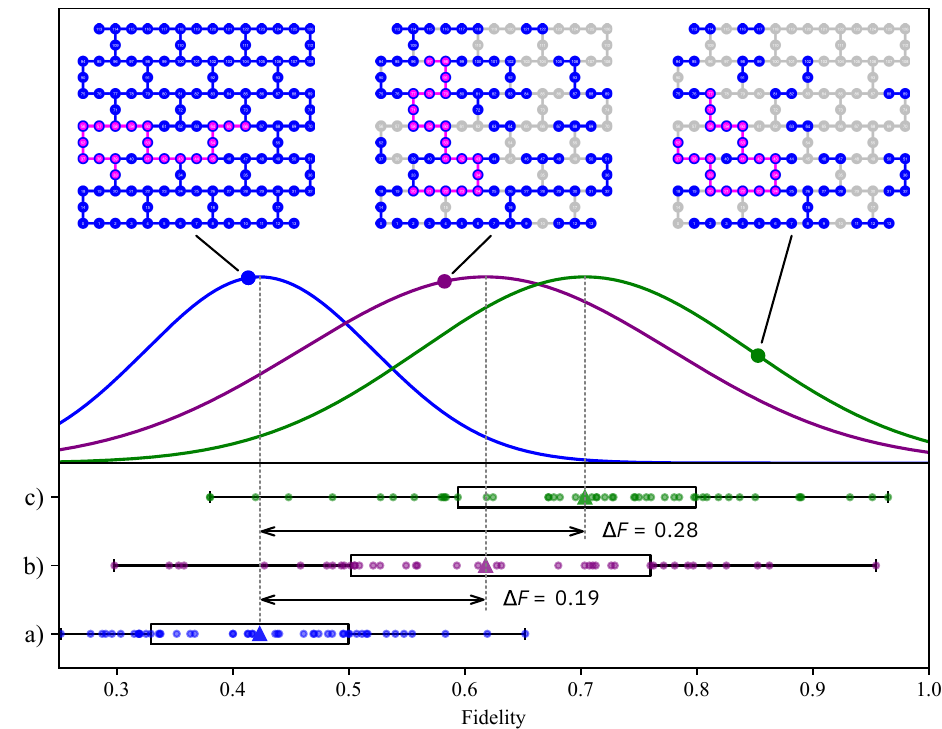}
     \caption{
     Normalized probability density distribution for random 20-qubit CNOT chains on \texttt{ibm\_quebec}, illustrating: a) the baseline fidelity across all qubits (in blue), b) the enhanced fidelity of partitions with CNOT error thresholds between 1.32\% and 1.39\% (in purple), and c) the enhanced fidelity with thresholds between 0.90\% and 1.16\% (in green). Insets showcase sample paths (magenta) within the pruned partitions (blue) or baseline, highlighting the remaining resources post-pruning.
 Data was collected from Dec. 23' to Jan. 24'.
}
    \label{fig:probadist20}
\end{figure}
\vspace{-0.1cm}
\subsection{Expected Gate Fidelity for 20-Qubit CNOT}
\vspace{-0.1cm}

Figure \ref{fig:probadist20} displays the normalized probability density distributions of gate fidelity for random 20-qubit CNOT chains executed within partitions generated by our method, with insets illustrating sample configurations and highlighting both utilized and pruned resources alongside the actual CNOT paths. We establish a baseline fidelity by utilizing all non-faulty qubits of the architecture, irrespective of their individual error characteristics, across 49 samples. To evaluate our method, we execute 249 random 20-qubit CNOT chains, adjusting the CNOT error threshold to confine their execution within progressively smaller partitions produced by our approach.
The analysis divides the results into two distinct sub-sampled groups from the original dataset: the 'moderately-pruned' group, with 49 samples derived from partitions at CNOT error thresholds between 1.32\% and 1.39\%, and the 'high-fidelity' group, comprising 41 samples from partitions with thresholds between 0.90\% and 1.16\%. These groups significantly outperform the baseline fidelity of 0.43, with mean fidelities of 0.62 and 0.71, respectively. The fidelity enhancements demonstrated by these sample distributions underscore the efficacy of our pruning method for 20-qubit CNOT chains. In the following section, we extend our examination to explore how our method impacts CNOT chains of varying lengths.
\vspace{-0.2cm}
\subsection{Expected Gate Fidelity for CNOT Chains of Varied Lengths}
\vspace{-0.1cm}
To comprehensively assess the effectiveness of our pruning method, we expand our investigation to varying the number of qubits in the CNOT chain. The experimental setup remains consistent with that presented in the previous section. Table \ref{tab:varyinglengths} presents the aggregated results of this extended evaluation. The table presents the baseline mean fidelity, standard deviation, and sample size for each chain length alongside the corresponding metrics achieved through our method. The final column, '$\Delta$ Mean,' highlights the percentage improvement in mean fidelity of our pruning method, underscoring its effectiveness across varying chain lengths. Those results demonstrate a pattern: as the length of the CNOT chains increases, the percentage improvement in gate fidelity also rises. This suggests that our pruning method is particularly beneficial for more complex quantum operations requiring longer entanglement chains. In summary, the extended benchmark reinforces the practical utility of our pruning method, demonstrating its capacity to significantly enhance gate fidelity across a broad spectrum of CNOT chain lengths. These findings affirm the method's potential to optimize quantum computational tasks of varying complexities, making it a valuable tool.

\begin{table}[]
    \centering
\scalebox{0.8}{\begin{tabular}{@{}cccccccc@{}}
\toprule
\multicolumn{1}{c}{\multirow{2}{*}{Qubits}} & \multicolumn{3}{c}{Baseline}                                                      & \multicolumn{3}{c}{Our Method}                                                     & \multicolumn{1}{c}{\multirow{2}{*}{$\Delta$ Mean}} \\ \cmidrule(lr){2-7}
\multicolumn{1}{c}{}                        & \multicolumn{1}{c}{Mean} & \multicolumn{1}{c}{Std. Dev.} & \multicolumn{1}{c}{N} & \multicolumn{1}{c}{Mean} & \multicolumn{1}{c}{Std. Dev.} & \multicolumn{1}{c}{N} & \multicolumn{1}{c}{}                               \\ \midrule
10                                            & 0.635                     & 0.146                         & 50                     & 0.719                     & 0.121                          & 203                    & \textbf{11.80\%}                                             \\
20                                            & 0.423                     & 0.098                         & 49                     & 0.646                     & 0.148                          & 269                    & \textbf{34.50\%}                                             \\
30                                            & 0.333                     & 0.065                         & 46                     & 0.581                     & 0.138                          & 213                    &\textbf{ 42.60\%}                                             \\
40                                            & 0.298                     & 0.037                         & 45                     & 0.569                     & 0.149                          & 169                    & \textbf{47.70\%}                                             \\
50                                            & 0.263                     & 0.016                         & 42                     & 0.549                     & 0.140                         & 121                    & \textbf{52.00\%}                                             \\ \bottomrule
\end{tabular}}
    \caption{Comparative Analysis of Gate Fidelity Across Varied CNOT Chain Lengths}
    \label{tab:varyinglengths}
\end{table}
\vspace{-0.2cm}
\subsection{Discussion}
A pivotal aspect of validating our methodology is the enforcement of an \texttt{initial\_layout}. This ensures that our benchmarks accurately reflect the characteristics of the entire partition rather than a biased subset of optimal resources. Without the use of \texttt{initial\_layout}, the transpiler defaults to optimizing the initial layout, which, given a particular calibration, would bias our results, offering an incomplete picture of the partition's overall performance. Future works should focus on integrating our method with existing noise-aware and RL based transpiler optimizations such as \cite{quetschlich_predicting_2023}\cite{wang_quest_2022}.
Our findings further substantiate the sensitivity of gate fidelity to the variance in noise levels across quantum resources. This insight led to a resource pruning approach, which, by selectively eliminating the most error-prone elements while retaining a large part of the architecture, improves gate fidelity by reducing the partition's error variance.

\vspace{-0.15cm}
\section{Conclusion}

We developed a method to generate high-fidelity partitions tailored for the highly-heterogeneous error characteristics of superconducting NISQ devices. 
By generating pruned partitions before the transpilation stage, we retain all transpiler functionality while  providing a guarantee that inadmissible resources have been pruned and thus will not be used.
Users can thus tailor their quantum application's size depending on pre-defined quantum resources performance requirements. This, in turn, can reduce the runtime needed to produce useful results for a given application, thereby reducing its cost.
For the case of random CNOT chains of length 50, our method demonstrated up to 52\% in gate fidelity over the non-pruned partition. Furthermore, we believe our method can extend the useful lifespan of existing quantum devices by pruning degraded resources and providing the user with a mode of operation---albeit with fewer qubits---that retains NISQ-level performances.
Our future work will focus on a deeper integration of our method with the transpiler, enabling the user to capture a different set of requirements for data and ancilla qubits.
\vspace{-0.15cm}
\section*{Acknowledgements}

We would like to thank the \textit{Plateforme d'Innovation Numérique et Quantique du Québec} (PINQ2) for the access to the machine \texttt{ibm\_quebec} and the computation time needed for this study.

\bibliographystyle{IEEEtran}
\bibliography{references}

\end{document}